\documentstyle[aps,epsf,prl,multicol]{revtex}

\def\Journal#1#2#3#4{{#1} {\bf #2}, #3 (#4)}

\def\CQG{\em Class. Quantum Grav.}

\def\PRD{\em Phys. Rev. D }
\def\GRG{\em Gen. Rel. Grav.}

\def\PRL{\em Phys. Rev. Lett.}

\def\MPL{\em Mod. Phys. Lett.} 
\def\PREP{\em Phys. Rept.}

\def\PLB{\em Phys. Lett. B } 

\def\NC{\em Nuovo Cimento} 
\def\EPL{\em Eurphys. Lett.} 
 
\newcommand{\bm}[1]{\mbox{\boldmath $#1$}} 
 
\def\g{{\rm g}} 
\def\espaitemps{({\cal V},\g)} 
\def\varietat{{\cal V}} 
\def\S{\Sigma} 
 
\def\doo{d \Omega^2_{S^3}} 
\def\trho{\tilde{\rho}}

\def\be{\begin{equation}} 
\def\ee{\end{equation}} 
\def\bea{\begin{eqnarray}} 
\def\eea{\end{eqnarray}} 
\def\bean{\begin{eqnarray*}} 
\def\eean{\end{eqnarray*}}

\newtheorem{result}{Result}[section] 
\begin{document} 
\setcounter{footnote}{2} 
\title{Signature Change on the Brane} 
\author{Marc Mars\footnotemark[1] , Jos\'e M. M. 
Senovilla\footnotemark[5] , and Ra\"ul Vera\footnotemark[4]\\  
\footnotemark[1] Albert Einstein Institut,  
Am M\"uhlenberg 1, D-14476 Golm, Germany. \\ 
\footnotemark[5] Departamento de F\'{\i}sica Te\'orica,  
Universidad del Pa\'{\i}s Vasco,
Apartado 644, 48080 Bilbao, Spain.\\ 
\footnotemark[4]  School of Mathematical Sciences,  
Queen Mary, University of London, 
Mile End Road, London E1 4NS, U.K. 
} 
 
\maketitle 
\begin{abstract} 
We explore the possibility of having a good description of classical 
signature change in the brane scenario. 
\end{abstract} 
 
PACS Numbers: 04.50.+h, 98.80.Cq, 11.10.Kk, 04.20.Gz.   
\begin{multicols}{2} 
 
\vspace{5mm} 
 
The aim of this letter is to show, in simple terms, that a natural  
scenario for the change of signature in the physical spacetime is  
provided by the brane-world models \cite{arkani,r-sI,r-sII} (see 
also \cite{barcelo,roberto,roberto2} for an exhaustive list of references) 
or, in general, by every  higher-dimensional theory \cite{hig-dim} 
which may contain   
domain walls and/or branes.

The main idea behind our proposal is that $d$-branes 
are nothing but {\it timelike} $(d+1)$-surfaces 
in a higher-dimensional spacetime (the bulk) \cite{Carter}.  
However, nothing prevents the  
possibility of having {\it perfectly regular} branes which change 
its character from (say) spacelike to timelike, or which are partly null,  
or even more complicated possibilities. The first  
case corresponds to a signature-changing brane.  
The  interesting property is that both the bulk and the brane can be regular 
everywhere even though the change of signature may  
appear as a dramatical event  
when seen from within the brane. 
Notice that the signature in the bulk is left unchanged, so that 
our work differs significantly from other recent studies  
\cite{bulk}. 
In our proposal, the study of the change  
of signature becomes the simple geometrical analysis of imbedded  
submanifolds in the bulk: a well-posed mathematical problem without 
pathologies.  
It is remarkable that many of the traditional {\it ad hoc  
assumptions} concerning signature change \cite{DEH} are shown to  
become pure {\it necessary conditions} in the brane case, which indirectly  
proves the plausibility of our idea and makes it worth exploring it, 
possibly sheding some light into the ``signature-change controversy'' 
\cite{DEH}.

Whether a signature change occurred in our effective  
spacetime is  
debatable,
and several independent works have considered this possibility  
\cite{ellissig,chsig}. From a classical viewpoint, a signature 
change may serve to  
avoid the singularities of general relativity \cite{sing},  
such as the big-bang, which might be replaced by a  
Euclidean region prior to the birth of time. 
Signature change has also been vindicated as an effective  
classical description of both the no-boundary proposal 
\cite{noboundaryprop} and the quantum tunneling \cite{vilenkintunnel} 
approach for the prescription of the Universe's 
wave function in quantum cosmology. 
In general, every process which can be studied by  
resorting to the ``imaginary time'', e.g. \cite{noboundaryprop}, 
can be also analyzed by means of change of the signature. 
All these possibilities could be naturally considered in our proposal.

As a matter of concreteness we will focus on the 
recent models based on a single 3-brane embedded into a 
five-dimensional Lorentzian manifold with a 
noncompact fifth dimension \cite{r-sII} (a more geometrically 
focused review of this model can be found in \cite{roberto}) 
and \cite{gogb1der,gogb2}. 
We can think of such a brane model as consisting of 
two Lorenztian regions  
joined together across corresponding smooth timelike boundaries.  
The matching between 
the two manifolds can be performed as long as the induced metrics 
on the two boundaries are isometric. 
The spacetime thus built 
is the bulk and the joining hypersurface $\S$ is the brane.  
The energy-momentum tensor on the brane can be calculated  
directly from the discontinuity $\left[K_{ab}\right]$  
of the second fundamental forms of $\S$ 
by using Israel's formula \cite{Israel}.  
Other kind of models with a compact fifth dimension, or based 
on two branes (e.g. \cite{r-sI}), 
or with a brane as the boundary of a single spacetime 
\cite{barcelo}, 
could be treated analogously. 
 
In order to describe signature-changing branes we only need to relax the 
condition that $\S$ is timelike, 
but then the usual matching conditions  
are no longer valid (in particular, the Israel formula) and 
the appropriate generalization must be used. 
Fortunately, this generalization was already developed in \cite{MS} 
in general relativity.  
The results carry over to any dimension with no essential change 
and can therefore be used to study signature-changing branes. So, let 
$\espaitemps$ be a 5-dimensional spacetime and $\S$ a 
smooth hypersurface $\S \subset \varietat$. The causal 
character of $\S$ is allowed to change along the hypersurface
More precisely, we assume that $\S$ contains three  
regions $\S^E$, 
$S$ and $\S^L$ where the hypersurface is spacelike, null, and timelike 
respectively: $\S^E$ will correspond to the 
Euclidean phase of the brane and $\S^L$ to  
the Lorentzian one, while the set $S$ (assumed, 
for definiteness, to have empty interior in $\S$) is
the signature-changing set. 
The brane $\S$ has a well-defined smooth normal 1-form $n_{\mu}$,  
which is timelike on $\S^E$, spacelike on $\S^L$ and becomes null at  
$S$. The induced metric on $\S$, or first fundamental 
form $h_{ab}$, is correspondingly positive-definite at $\S^E$,  
Lorentzian at $\S^L$, and degenerate at $S$ \cite{MS}. As is  
known, one degeneration vector at $S$ is precisely $\vec n$  
(index upstairs), which is tangent to $\S$ at $S$ \cite{MS}. A simple 
but important  consequence of the construction is
the following result \cite{largo}.
\begin{result} 
The signature-changing set $S$ is a smooth 
spacelike three-dimensional surface. The induced metric 
$h_{ab}$ of $\S$ has $\vec{n}|_{S}$ as unique degeneration  
direction at $S$. 
\end{result} 
In plain words, by choosing the reference system appropriately, this  
means that the signature change takes place at an {\it instant of time}. 
Since this property is desirable, it has always been implicitly 
{\it assumed}, 
However, in a pure 4-dimensional spacetime, not imbedded in a bulk, there 
exist {\it many} other possibilities. Interestingly, 
this becomes now a prediction, 
providing a first clear  
example of how the brane scenario can lead to strong limitations on 
the allowed possibilities thereby proving that the traditional 
ad hoc assumptions are justified and natural. 
 
But, can we actually produce a sensible signature-changing brane?  
To answer this, we have examined the traditional ways  
of building explicit branes.  The simplest, and  
most frequently used, method to 
construct them is to cut a spacetime across a timelike hypersurface 
and join it to an identical copy of itself across the boundary. 
The resulting bulk has a $Z_{2}$-symmetry with respect to the brane.  
It is natural to ask whether a similar construction can  
produce signature-changing branes.  
\begin{result} 
\label{res:2} 
{\em \cite{largo}} 
It is impossible to join two identical copies of a spacetime 
with signature-changing boundary $\S$, across $\S$, 
to produce a bulk with continuous metric. 
\end{result} 
Hence, the $Z_{2}$ mirror symmetry is incompatible 
with a signature-changing brane. Therefore, for signature-changing branes, 
more sophisticated 
constructions are necessary, 
such as gluing two different regions 
of the same spacetime, or two different 
spacetimes across appropriate hypersurfaces. 
Another consequence is that this result may select  
the proper construction for the Riemann tensor of a manifold with 
boundary \cite{barcelo}, because the $Z_2$-symmetry 
used in one of the two procedures in \cite{barcelo} cannot be invoked 
when the boundary changes its character.

Another standard procedure is the use of 
umbilical hypersurfaces, i.e. those for which the second fundamental form is  
proportional to the first one everywhere. 
This immediately implies  
that the energy-momentum 
tensor on the brane is of cosmological constant type. However 
\begin{result} 
\label{res:3} 
{\em \cite{largo}} 
A smooth umbilical hypersurface must have constant 
signature. Moreover, if $[K_{ab}] = F h_{ab} \neq 0$ on a brane 
$\S$, then its signature must remain constant. 
\end{result} 
Therefore, everywhere umbilical branes cannot undergo  
a change of signature. A physical consequence is that  
signature-changing branes cannot have a $\Lambda$-term 
energy-momentum tensor {\it everywhere}. Hence, some 
fields must become excited at least  
near the signature-changing set $S$. This seems 
to indicate the existence of some 
dynamical quantum processes for the fields present,  
responsible for the eventual 
change of signature.  Nevertheless, our treatment is intended to describe 
a pure classical limit of any quantum mechanism leading to the 
signature change, and it has enough freedom to allow for specific models 
in this direction.

The above Results show essential differences between signature-changing 
and standard timelike branes.  
In the sequel, we show the existence of 
signature-changing branes with several desirable features 
by presenting an explicit example \cite{largo}.

Because of its importance in the Randall-Sundrum models 
\cite{r-sI,r-sII} 
and as is customary in brane and string works, the bulk will be taken 
to be anti-de Sitter spacetime, $AdS_{5}$, which in  
adequate coordinates has the metric 
\be 
ds^2=-(1+\lambda^2 \rho^2)dt^2+(1+\lambda^2 \rho^2)^{-1}d\rho^2+\rho^2\doo, 
\label{AdS} 
\ee 
where $\doo$ is the round metric of $S^3$,  
$\rho>0$,
$\lambda>0$ is a constant. 
The cosmological constant is 
$\Lambda =-6 \lambda^2$ (flat bulk as
$\lambda \rightarrow 0$). For the sake of simplicity we will  
only consider the spherically symmetric hypersurfaces $\S$ described 
by $F(t,\rho)=0$, or equivalently in parametric form 
(ignoring the angular part) by $t(\xi)$, $\rho(\xi)$,  
where $\xi$ is the parameter. With this  
assumption, the first fundamental form of $\S$ is 
\be 
ds^2|_{\S}= N (\xi)d\xi^2 +a^2(\xi)\doo,\label{RW} 
\ee 
where $a(\xi)\equiv \rho(\xi)$,  
$N \equiv -n_{\mu}n^{\mu}=(1+\lambda^2 a^2)^{-1}\dot{a}^2- 
(1+\lambda^2 a^2)\dot{t}^2$ and overdot means $d/d\xi$. 
The change of signature corresponds obviously to a change in the sign 
of $N(\xi)$. Expression (\ref{RW}) has two desirable features:  
the Lorentzian part of $\S$ describes a standard (closed)  
Robertson-Walker (RW) cosmological  
model; and the change of signature happens everywhere at some instant  
of cosmological time, thus replacing the universal big bang. 
The model still has one free function of $\xi$ 
which can be chosen according  
the particular situation being tackled.  
 
\begin{figure}
\centering 
\mbox{}
\epsfxsize=4cm\epsfbox{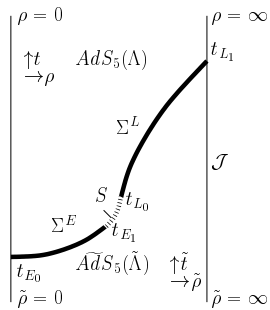}
\caption{
Conformal diagram for $AdS_{5}$ and $\widetilde{AdS}_5$ 
matched along the changing signature hypersurface $\S$, represented 
by the curve going from $t_{E_0}$ to $t_{L_1}$ 
The bulk is the product of this diagram with $S^3$. 
As usual, null lines are at 45$^o$. 
The umbilical regions $t_{E_0}< t < t_{E_1}$ in $\S^E$ and 
$t_{L_0}< t <t_{L_1}$ in $\S^L$ are denoted by solid lines, 
whereas the non-umbilical intermediate region is denoted by a dotted line. 
} 
\label{fig:2} 
\end{figure}

We can now proceed to the construction of the brane. 
Due to Result \ref{res:2}, we cannot use the standard 
procedure of gluing two copies of $AdS_{5}$ across 
the boundary $\S$. However, we can still keep $AdS_{5}$ as our global 
bulk by taking {\it another different} $\widetilde{AdS}_{5}$ with a 
{\it different} cosmological constant $\tilde{\Lambda}=-6 \tilde{\lambda}^2$ 
and line-element 
\begin{eqnarray} 
d\tilde{s}^2=-(1+\tilde{\lambda}^2 \trho^2)d\tilde{t}^2+ 
(1+\tilde{\lambda}^2 \trho^2)^{-1}d\trho^2+\trho^2\doo, 
\end{eqnarray} 
and a new spherically symmetric hypersurface $\tilde{\S}$ given in 
parametric form by $\trho (\xi)$ and $\tilde{t}(\xi)$. 
A necessary requisite in  order to build a well-defined bulk by 
pasting $\S$ with $\tilde{\S}$ is that the corresponding first  
fundamental forms (\ref{RW}) 
of $\S$ in $AdS_{5}$ and of $\tilde{\S}$ in $\widetilde{AdS}_{5}$ be  
isometric.  
This fixes $\tilde{\S}$ completely (except for isometries)  
as the solution of 
$\trho(\xi)=\rho(\xi)\equiv a(\xi)$ and of the differential equation 
\begin{eqnarray} 
{\dot{\tilde{t}}}^{2} = (1+\tilde{\lambda}^2a^2)^{-2}[\dot{a}^2-N(\xi) 
(1+\tilde{\lambda}^2a^2)]. 
\label{ODE} 
\end{eqnarray} 
By using the results in \cite{MS} we can compute the 
energy-momentum tensor of the resulting bulk which, as in the standard 
timelike case, has a distributional part 
$T_{\mu\nu} |_{\S} = \delta \cdot \tau_{\mu\nu}$, where $\delta$ is 
a typical scalar distribution with support on the brane. Some care is 
needed here, because the definition of $\delta$ requires a 
choice of volume form \cite{MS}, which is canonical  
when the hypersurface is everywhere non-null, 
but not for a signature-changing $\S$. Nevertheless, 
$\delta \cdot \tau_{\mu\nu}$ is independent of this choice \cite{MS} 
(but $\tau_{\mu\nu}$ {\it is} choice dependent!). 
Selecting the volume 4-form  
$\bm{\eta} = a^3 d \xi \wedge {\bm\eta}_{S^3}$ 
on $\S$, with $\bm{\eta}_{S^3}$ the standard measure in $S^3$, 
and for a matching as in figure \ref{fig:2} 
(where $\dot{t}>0$ along $\Sigma$ \cite{largo}) 
the explicit expression for $\tau_{\mu\nu} dx^{\mu} dx^{\nu}$ reads 
\begin{eqnarray} 
- \frac{\varrho}N [ \dot{t}(1+\lambda^2 a^2) dt - 
\dot{a}(1+\lambda^2 a^2)^{-1} d\rho ]^2 + p\, a^2 \doo, 
\label{EM} 
\end{eqnarray} 
\begin{eqnarray} 
\varrho = \frac{3}{a {\kappa_5}^2}[ 
\sqrt{\dot{a}^2-N(1+\lambda^2)}- 
\sqrt{\dot{a}^2-N(1+\tilde{\lambda}^2)} 
] 
\label{eq:rho} 
\end{eqnarray} 
where $\kappa_5$ is the coupling constant corresponding to the Einstein 
equations in 5 dimensions, 
and $p$ can be obtained from the following conservation equation 
\begin{eqnarray} 
\dot{\varrho} - \varrho \frac{\dot{ N}}{2 N} +  
\frac{3\dot{a}}{a} \left (\varrho + p \right ) =0. 
\label{eq:cons} 
\end{eqnarray} 
A simple analysis shows that $\tau_{\mu\nu}$ and  
$\delta \cdot \tau_{\mu\nu}$ are 
regular everywhere on $\S$. 
Equation (\ref{eq:cons}) takes the usual RW form 
for the variables $\varrho\, |N|^{-1/2}$ and $p\, |N|^{-1/2}$. 
Thus, at points not in $S$, 
$n_\mu$ can be normalized 
and the usual conservation equation in $\S^L$ {\em and} 
$\S^E$ is recovered \cite{largo}. 
It must be stressed here that $\varrho$ and $p$ 
in (\ref{EM})-(\ref{eq:rho}) are 
naturally defined as eigenvalues of $\tau_{\mu\nu}$, 
without invoking ad hoc assumptions,  
in contrast with the definition of $\varrho$ given in earlier 
works \cite{ellissig} which has opposite sign in $\S^E$. 
 
\noindent

The global bulk thus defined is then constituted by two different regions of  
$AdS_{5}$-type, separated by a brane which, if desired, can change  
signature. Timelike branes separating two $AdS_5$ bulks  
with different cosmological constants have been already studied, 
\cite{gogb1der}, and are included in our treatment. 
The two $AdS_{5}$ regions  
may be interpreted as two fundamental states with different vacuum  
energies which can live together precisely due to the existence of the 
brane.
Models describing jumps of the cosmological 
constant have been presented in
different
contexts, mainly in order 
to explain its small present day value (see \cite{Feng} and references 
therein).
In the case of signature-changing $\S$, 
if the Lorentzian part of the brane is connected  
one can easily see 
that there must always exist a time $t_{E_0}$ in the coordinates 
we are using such that the bulk is $\widetilde{AdS}_{5}$ for {\it all 
times} before $t_{E_0}$, see Fig. \ref{fig:2}.  
Thus, we can think of $\widetilde{AdS}_{5}$ 
as the original bulk, which may represent a false vacuum. 
This vacuum would undergo a phase transition (similar to that of 
standard inflation, for instance) which occurs, as usual, in an 
acausal way, so that it can be modeled with the spacelike part $\S^E$. 
In a region near $S$, and due probably to the matter fields present 
and their properties,  
$\S^E$ undergoes an internal process of signature change 
and it smoothly changes to the Lorentzian part $\S^L$. Then, there is an 
epoch (up to the time $t_{L_1}$, see Fig. \ref{fig:2}) in which the two 
vacua co-live separated in fact by a timelike brane. This part $\S^L$ 
of the brane would be our 4-dimensional world. Eventually, for all 
times after $t_{L_1}$, the bulk becomes $AdS_{5}$ with the constant 
$\Lambda$. Notice further that both the bulk and the brane are 
regular everywhere. In the brane, there appears a 
very distinguished instant of time, given by $S$, 
and a transition region around $S$ (one part belonging to 
$\S^E$ and another part to $\S^L$), 
which are quite remarkable {\it from the inner 
point of view of the brane}. They would correspond to the big-bang 
`singularity', to the pre-big-bang  Euclidean phase, and to the very 
early universe (possibly with an  inflationary era), respectively. 

Of course, any phase transition  
takes some (very small but finite) time, and thus the  
hypersurface description used here for $\S^E$  
is an effective one. In our opinion, this is yet another  
positive property of our proposal, because it makes the explicit  
models {\it theoretically testable}, in the following sense. 
There must be a relation  
between the thickness of $\S^L$---which is its {\it spatial  
extension}---, 
and that of $\S^E$---which is its {\it temporal duration}--- being 
both part of the same brane. For instance, in some brane 
scenarios \cite{arkani} the thickness of $\S^L$  
is of the order of
the electroweak scale $m_{EW}\sim 1$TeV$\sim 10^{-16}$mm. 
This gives an estimation for the thickness of $\S^E$ of around 
$10^{-28}-10^{-29}$s and this should be in agreement with the time 
scales for the phase transitions in any microscopic proposal to 
describe the decay from $\tilde{\Lambda}$ to $\Lambda$.  
Obviously a similar restriction would happen if a different brane 
model, and hence a different thickness
of $\S^L$, is considered.
Other thickness estimations would come from
the length scales of graviton trapping 
\cite{r-sII}, given by $(-6/\Lambda)^{1/2}$ \cite{barcelo}, and also 
either from distances between branes or radius
of compact extra dimensions ($<$ 1mm for experimental reasons).

There are very many possibilities to construct explicit models of the  
type we are considering. The free function $a(\xi)$ 
can be determined once the matter on the brane is 
chosen. For simplicity, let us consider the case of a 
scalar field, assumed to have an unstable constant value 
in most of $\S^E$, and to finally settle down to another 
stable constant value in most of $\S^L$. Thus, the brane has a cosmological 
constant type energy-momentum in some large portions of both 
$\S^E$ and $\S^L$. Only around 
the signature-changing set $S$ the scalar field becomes dynamical.  
It is easy to  
see from (\ref{EM})-(\ref{eq:cons}) that $\tau_{\mu\nu}$ will take 
the form of a cosmological constant 
energy-momentum tensor if 
$\dot{a}^2-N [a^2\alpha^2 \lambda^2/(\alpha^2-1)+1]= 0$, 
where $\alpha>0$ is a constant. Its general solution leads to 
the following implicit form of $\S$:
\be 
F(t,\rho)= 
\alpha\sin\left\{\lambda(t-t_{L_1})\right\}\sqrt{1+\lambda^2 \rho^2}-1=0 
\label{Sigma} 
\ee 
where $t_{L_1}$ is a constant.
The family (\ref{Sigma}) corresponds to the spherically 
symmetric umbilical hypersurfaces in $AdS_5$, and their scalar 
curvature is given by ${^{(4)}}R=12 \lambda^2 \alpha^2/(1-\alpha^2)$.  
These $\S$ are spacelike 
for $\alpha >1$, timelike for $0<\alpha<1$ and null for $\alpha=1$. From 
our assumptions, the brane will be umbilical everywhere except 
for a region around $S$. 
Notice, however, that this 
transition region can be made {\it as small as desired}.  
We choose to describe the entire brane by keeping the functional 
form (\ref{Sigma}) and letting $\alpha$ become a function of $\xi$, 
which can be taken as any smooth function of $t$ ($\dot{t} >0$) in the interval 
$(t_{E_0},t_{L_1})$ with the following properties: 
$\alpha = \alpha_1 > 1 $ for $t_{E_0}< t < t_{E_1}$, 
$\alpha= \alpha_2 <1$ for $t_{E_1}<t_{L_0}< t <t_{L_1}$ and, 
in the intermediate region $t_{E_1} < t <  t_{L_0}$, $\alpha(t)$ is an 
interpolating function between the two constants $\alpha_1$ and $\alpha_2$. 
Observe 
that $t_{E_1}<t_{E_0}+\pi/(2\lambda)$ and $t_{L_0}>t_{L_1}-\pi/(2\lambda)$, 
so that the only requirement is that $\alpha_1^2+\mu(1-\alpha_1^2)>0$, 
where we have set $\mu\equiv \tilde{\lambda}/\lambda$. 
The change of signature must necessarily happen in the transition 
region $t_{E_1} < t <  t_{L_0}$.  
 
The form of the brane as seen form $\widetilde{AdS}_5$ can be 
found from (\ref{ODE}). It can be easily proven that the hypersurface 
in $\widetilde{AdS}_5$  
is also umbilical in $\S^E$ and $\S^L$. Hence, it  must take 
the form (\ref{Sigma}) where $\rho,t,t_{L_1} \rightarrow \trho, \tilde{t}, 
\tilde{t}_{L_1}$. The corresponding constant $\tilde{\alpha}$ can be found 
from the matching conditions to  be 
$\alpha /\sqrt{ \alpha^2+ \mu^2 (1 - \alpha^2 )}$.  
Since the hypersurface is nowhere null in the umbilical 
regions, we can take a unit  
$n_{\mu}$ in order to 
define the distribution $\delta$ at $\S^E$ and $\S^L$. 
With this choice, $\tau_{\mu\nu}$ takes the form 
\begin{eqnarray} 
\left.\tau_{\mu\nu}\right|_{(t_{E_0},t_{E_1})}=&&
\frac{3 \lambda}{\sqrt{\alpha_1^2 -1}}
[\sqrt{\mu^2 \left (1- \alpha_1^2 \right ) + \alpha_1^2} -1 ] \nonumber\\ 
&&\times \left ( g_{\mu\nu} + n_{\mu} n_{\nu} \right ),  
\nonumber \\ 
\left.\tau_{\mu\nu}\right|_{(t_{L_0},t_{L_1})}=&&
\frac{3 \lambda}{\sqrt{1 - \alpha_2^2}}
[\sqrt{\mu^2 \left (1- \alpha_2^2 \right ) + \alpha_2^2} -1 ] \nonumber\\ 
&&\times \left ( g_{\mu\nu} - n_{\mu} n_{\nu} \right ). 
\label{Lamb} 
\end{eqnarray} 
These expressions show that the tension on the umbilical region of $\S^L$ 
is positive if and only  
if $\tilde{\lambda} < \lambda$. This has a nice physical 
interpretation because  
the energy-density of the  
original bulk $\widetilde{AdS}_5$ is less negative  
than the energy-density of the final bulk $AdS_{5}$, in accordance 
with the possibility that $AdS_5$ is more stable than 
$\widetilde{AdS}_5$. 
Furthermore, we find from 
(\ref{Lamb}) that, when $\tilde{\lambda} < \lambda$, the energy-density 
on the umbilical part of $\S^E$  
measured by {\it any} timelike observer is also positive. Again this 
is physically reasonable. 
 
Of course, many other possibilities 
are allowed and, for any type of energy-momentum tensor, 
(\ref{EM}) can be solved to get $\S$.  
Hence, any closed RW brane can be modeled  
by our construction from times not too close to $S$.

We are grateful to R. Emparan for many enlightening  
conversations. We thank financial support 
from the Universidad del Pa\'{\i}s Vasco  
project No. UPV 172.310-G02/99. R.V. thanks the 
Spanish SEUID
Grant No. EX99 52155527. M.M. and J.M.M.S. wish to thank the Albert Einstein
Institut,  and the Institut f\"ur Theoretische
Physik, University of Vienna, respectively,
for kind hospitality.

\end{multicols} 
 
\end{document}